# Cosmology and the Origin of the Universe:
# Historical and Conceptual Perspectives

Helge Kragh*

**Abstract**: From a modern perspective cosmology is a historical science in so far that it deals with the development of the universe since its origin some 14 billion years ago. The origin itself may not be subject to scientific analysis and explanation. Nonetheless, there are theories that claim to explain the ultimate origin or "creation" of the universe. As shown by the history of cosmological thought, the very concept of "origin" is problematic and can be understood in different ways. While it is normally understood as a temporal concept, cosmic origin is not temporal by necessity. The universe can be assigned an origin even though it has no definite age. In order to clarify the question a view of earlier ideas will be helpful, these ideas coming not only from astronomy but also from philosophy and theology.

## 1. Introduction: concepts of origin

In a lecture of 1993 the famous British astrophysicist and cosmologist, Fred Hoyle, recommended that "Whenever the word 'origin' is used, disbelieve everything you are told." He added, "The biggest pig in a poke where origins are concerned is that of the whole universe."[1] Whatever the opinion of Hoyle the concept of origin is generally difficult and often tricky, for other reasons because the term is used in different contexts and with different meanings. According to the Oxford English Dictionary, origin means "the act or fact of beginning, or of springing from something; beginning of existence with reference to source or cause; rise or first manifestation."[2] Several other, more or less synonymous words are in use, such as start, genesis (or generation), birth, dawning, emergence, and creation. However, in reality the words cover a wide spectrum of meanings and are thus not proper synonyms.

---

* Niels Bohr Institute, University of Copenhagen, Denmark. E-mail: helge.kragh@nbi.ku.dk. This paper is an English version of a chapter to be published later in 2017 in a book in Spanish, titled *Orígenes* and edited by Ángel Díaz de Rada at the Open University in Spain (UNED).
[1] Fred Hoyle, *The Origin of the Universe and the Origin of Religion* (London: Moyer Bell, 1993), p. 18.

Typically, when we ask "what is the origin of *X*" we inquire about how *X* was once formed or came into being. A satisfactory answer is supposed to relate to certain states of the past from which *X* emerged. When we say that a person's country of origin is Algeria we mean that he or she was born and raised in Algeria. As a more relevant example, consider the origin of the Earth, a question which for long was a puzzle to geologists and astronomers but is now well (if not completely) understood. According to the consensus theory the Earth was formed approximately 4.5 billion years ago from gravitationally caused accretion of matter particles in the solar nebula. If this theory turns out to be correct we will have scientifically valid knowledge concerning the origin of the Earth. In some cases the origin of *X* is something which happened almost momentarily but in other cases it is a slow and gradual process to which no definite time in the past can be assigned. Given that the Earth's crust solidified $4.54 \pm 0.05$ billion years ago, the origin of the Earth took place quickly, at least on a geological time scale.

The case of the origin of the biological species, as first fully addressed in Charles Darwin's famous *The Origin of Species* from 1859, is in this regard quite different. Zoologists can explain how and approximately when the mammoth evolved from earlier elephants, but in this and some other cases origin means essentially evolution. Indeed, to account for the origin of *X* often means to come up with a history of how *X* developed from ancestral forms. The two cases have in common that there was a time when *X* did not exist. For 6 billion years ago there was no Earth, and for 10 million years ago there were no mammoths. We will not normally wonder about the origin of something which has always existed and to which a first instance cannot be ascribed (but see Section 2). Likewise, we will not normally wonder about the origin of something which has never existed – we may be interested in the origin of the belief in unicorns but not in the origin of unicorns.

In many cases *X* refers to an object or a class of objects, but it may also be a phenomenon. What is the origin of the enigmatic colours of the aurora borealis? In this case scientists will point to mechanisms in the atoms in the upper strata of the atmosphere, at a height of approximately 100 km, which are responsible for the wavelengths identified in the spectrum of the aurora. Similarly, to explain the origin of the present phenomenon of global warming they will examine models of how solar heat behaves in an atmosphere containing carbon dioxide and other greenhouse gases. In none of these cases does the concept of origin have a direct temporal meaning as it primarily relates to a causal mechanism. Nonetheless, since the phenomena are effects of certain causes, and causes always precede the effects, there is implicitly a temporal link.

---

[2] http://www.oed.com (accessed 10/08/2026).

When we consider the so-called origin of the universe, things become much more difficult. In the mentioned examples the origin of $X$ is taken to refer to either an ancestral state or something from which the existence of $X$ can be derived either causally or in some other way. Consider a radioactive decay where a certain atom $Y$ transmutes into another atom $X$; since the $X$ atom did not exist before the decay we may say that it was created. The decay process is governed by the probabilistic laws of quantum mechanics which means that no cause can be given why $X$ was created at a definite time. Nevertheless, although a sufficient cause cannot be given, the decay depends on necessary causes. For example, if there were no atoms of type $Y$ the creation of $X$ would not happen. It depends on a prior state, which distinguishes it from the case of the creation or origin of the big bang universe.

In most cases it makes sense to say that $Y$ was the cause of $X$; by specifying $Y$ we can account for the origin of $X$ but only by making the perhaps obvious assumption that $Y$ and $X$ are parts of the universe. Given the standard definition of the universe we cannot speak in the same manner if $X$ = universe. It is of crucial importance to be aware that the two statements "the Earth is 4.5 billion years ago" and "the universe is 13.7 billion years ago" are entirely different. If the universe has an absolute origin it presumably means that it came into being as a result of something either before or outside the universe. None of the options seem to make much sense from a scientific point of view and perhaps not even from a logical point of view. Sometimes the question of the origin of life is considered to be of a status similar to the one of the origin of the universe. But this is a mistake, I think. Although we do not know the answer to the first question it can presumably be answered in the traditional way, that is, to find out when and how primitive life forms first evolved from complex organic but abiotic molecules.

To discuss the question of the origin of the universe in either an evolutionary or an absolute version I shall adopt a historical approach.[3] Despite the enormous progress cosmology has made during the last century the origin problem has not come substantially closer to a solution. From a philosophical rather than scientific perspective it was much discussed during the previous centuries, and some of the arguments entering the discussions are still worth to recall. Precisely because they were not limited to a modern scientific perspective they sometimes included viewpoints concerning the origin of the universe of conceptual originality and interest.

## 2. Ancient and medieval periods

The question of the origin of the universe has not always constituted a significant part of cosmological thought. Although there have been periods in which it was considered important, such as is obviously the case today, through most of history the question failed to attract much attention or was even considered outside the scientific study of the cosmos. Yet the very earliest concerns with the heavens were in the form of *cosmogonies* rather than cosmologies, meaning that they were attempts to understand how the present world or universe had come into existence. According to the cosmogonies of the ancient Egyptian and Mesopotamian civilizations, the universe was created as a dynamic entity which gave rise to everything existing, including humans and gods.[4] The stories that the learned ancients told about the origin and evolution of the universe were not scientific but mythological. Because of the crucial role played by the gods the stories were *theogonies* as much as they were cosmogonies. The two genres were indistinguishable.

A common feature to be found in both of the great river cultures was the belief that the creative process started with an undifferentiated watery chaos which the gods subsequently separated into two or more separate realms, thereby creating Earth and the heaven as distinct bodies. The original state of darkness and lifeless uniformity was posited but not explained. In any case, the primary concern of the early cosmogonies was not to describe the universe but to account for the existence of its inhabitants in the form of gods, humans and everything else. This is the kind of story we meet in the Mesopotamian creation myth *Enuma Elish*, dating from about 1500 BC, and also in the much later *Theogony* written by the Greek poet Hesiod. A different variant of the creation theme appears in the Jewish cosmogony as described in the very beginning of Genesis:

> In the beginning, when God created the universe, the Earth was formless and desolate. "Let there be light – and light appeared." … Then He separated the light from the darkness … Then God commanded, "Let the water below the sky come together in one place, so that the land will appear" – and it was done. He named the land "Earth" and the water which had come together he named "Sea."

---

[3] Modern histories of cosmology dealing with the subject from the oldest time to the present include John North, *Cosmos: An Illustrated History of Astronomy and Cosmology* (Chicago: University of Chicago Press, 2008) and Helge Kragh, *Conceptions of Cosmos. From Myths to the Accelerating Universe: A History of Cosmology* (Oxford: Oxford University Press, 2007).

[4] See, for example, Carmen Blacker and Michael Loewe, eds., *Ancient Cosmologies* (London: George Allen & Unwin, 1975) and chapters in Norriss S. Hetherington, ed., *Encyclopedia of Cosmology: Historical, Philosophical, and Scientific Foundations of Modern Cosmology* (New York: Garland Publishing, 1993).

The crucial novelty, apart from the introduction of a single almighty and non-created God, is the absence of a primordial state. So, from what did God create the universe? The text does not say that he created it out of nothingness, but in later Christianity it became a dogma that this is what happened. First there was nothing and then, because God wanted it, there was something. *Creatio ex nihilo*.[5] As early as the fifth century AD the church father St. Augustine pointed out that God's creation of the world included time itself. "The world was made, not in time, but simultaneously with time," he wrote.[6] There was no time before the universe.

With the rise of Greek natural philosophy and its gradual transformation into science, astronomy emerged as a mathematical discipline focused on observations of the planetary system; on the other hand, speculations about the origin of the world declined drastically.[7] In the long period between Plato and

---

[5] God's creation of the world *ex nihilo* dates from the second century but was only made an official doctrine of the Catholic Church at the Fourth Lateran Council in 1215. Experts disagree of whether or not the idea is implied by Genesis or other parts of the Bible. For opposing views, see Gerhard May, *Creatio ex Nihilo: The Doctrine of "Creation Out of Nothing" in Early Christian Thought* (Edinburgh: T & T Clark, 1994) and Paul Copan and William L. Craig, *Creation Out of Nothing: A Biblical, Philosophical, and Scientific Exploration* (Grand Rapids: Baker Academic, 2004).

[6] Vernon Bourke, ed., *The Essential Augustine* (Indianapolis: Hackett Publishing Company, 1974), p. 109.

[7] On aspects of Greek cosmological thought, see Samuel Sambursky, *The Physical World of the Greeks* (London: Routledge, 1963), Richard D. McKirahan, *Philosophy Before Socrates*

Ptolemy cosmogony, in the strict meaning of the term, practically came to a halt. However, questions concerning the origin of the world and its subsequent evolution continued being addressed by natural philosophers outside the mainstream Aristotelian tradition. One of them was the Roman poet Titus Lucretius Carus, an advocate of the much earlier atomistic philosophy going back to Democritus and Epicurus.

In a famous text composed about 50 BC called *De Rerum Natura* (On the Nature of Things), Lucretius described the atomists' infinite universe solely made up of atoms in incessant motion. This universe was of finite age and "there will be an end to the heaven and the Earth."[8] Rather than basing his argument on some mythological scenario Lucretius called attention to the shortness of human history. Apparently unable to conceive a world without humans (or without poets), he asked: "If there was no origin of the heavens and Earth from generation, and if they existed from all eternity, how is it that other poets, before the time of the Theban war, and the destruction of Troy, have not also sung of other exploits of the inhabitants of Earth?" Lucretius suggested that "the whole of the world is of comparatively modern date, and recent in its origin." He and a few other philosophers of the atomistic and Stoic schools argued that "The walls of the great world, being assailed around, shall suffer decay, and fall into mouldering ruins." From this they concluded that the world cannot be eternal in the past but must have had a beginning in time.[9]

The philosophical cosmology of Lucretius, including its associated cosmogony, differed completely from the far more influential ideas of Aristotle expounded in his *De Caelo* (On the Heavens) and other works. Aristotle's cosmos was in a steady state in so far that it was eternal, and local non-circular changes were restricted to the sublunary world. He argued that the universe as a whole, apart from being unique (no other universes), was spatially finite but temporally infinite in both

---

(Indianapolis: Hackett Publishing Company, 1994), and M. R. Wright, *Cosmology in Antiquity* (London: Routledge, 1995).

[8]  Lucretius, *On the Nature of Things* (Amherst, NY: Prometheus Books, 1997), quotations from pp. 45-46, 93, and 205.

[9]  Versions of this kind of argument against the eternity of the world can be found much earlier. The Stoic philosopher Zeno of Citium developed it on the basis of the observed erosion processes on the Earth's surface. See Gad Freudenthal, "Chemical foundations for cosmological ideas: Ibn Sina on the geology of an eternal world," pp. 47-73 in Sabetai Unguru, ed., *Physics, Cosmology and Astronomy 1300-1700: Tensions and Accommodation* (Dordrecht: Kluwer, 1991).

directions. In other words, it was eternal and hence uncreated as well as indestructible.

Suppose, Aristotle said in *De Caelo*, that the world was formed out of some pre-existing and unchanging elements. "Then if their condition was always so … the world could never have come into being. And if the world did come into being, then, clearly, their condition must have been capable of change and not eternal."[10] He further referred to what in cosmological thought is known as the "why not sooner?" argument, namely: If the universe came into being a finite time ago, what reason could there possibly be for just this time rather than some other time? In the words of Aristotle: "Why, after an infinity of not being, was it generated, at one moment rather than another? If there is no reason and the moments are infinite in number, it is clear that a generated or destructible thing existed for an infinite time." Augustine would later counter that the universe could not possibly have come into existence at an earlier time since there was no time before the universe began.[11] Aristotle summarized his position as follows: "The heaven as a whole neither came into being nor admits of destruction, … but is one and eternal, with no end or beginning." Aristotle's assumptions about a finite and eternal cosmos were not generally accepted in the Greek-Roman culture, but later they came to dominate the medieval world view. There was one exception though, namely the controversial and most un-Christian claim of the universe being past eternal.

The Christian universe was divinely created out of nothing and hence of finite age, in sharp contradiction to what Aristotle had taught. As early as the sixth century the Christian philosopher Johannes Philoponus developed a series of rational arguments based on the concept of infinity against Aristotle's heresy. Let us assume, Philoponus said, that the world had always existed and been populated with humans. In that case, there would have existed an infinity of humans up to the time of Socrates. But, he went on, "there will have to be added to it the individuals who came into existence between Socrates and the present, so there will be something greater than infinity, which is impossible."[12] Another variant of his argument related to the periods of revolution of the planets and stars. If Saturn had revolved infinitely many times, Jupiter would have performed three times as many revolutions and the stars many more times as many. This he thought was not only incredible but strictly impossible. "Thus necessarily the revolution of the heavenly bodies [and hence the universe itself] must have a beginning."

---

[10]  Quotations from Jonathan Barnes, ed., *The Complete Works of Aristotle*, vol. 1 (Princeton: Princeton University Press, 1984), pp. 461-470.

[11]  The "why not sooner?" argument against a cosmic beginning can be found in the pre-Socratic philosopher Parmenides and was later discussed by Leibniz and Kant. See Brian Leftow, "Why didn't God create the world sooner?" *Religious Studies* **27** (1991), 157-172.

Much later, when Aristotle's philosophical system had been rediscovered and to a large part incorporated in the medieval-Christian world view, the problem of cosmic origin remained controversial. It caused the great theologian Thomas of Aquinas to re-examine the idea of creation. Could God have created an eternal universe? Creation and eternity may appear to be mutually exclusive concepts, but Aquinas pointed out that since God is a non-temporal being he did not need to precede his effects in time. God did not transform "nothing" into something, he causes things to exist continually in the sense that if they were left to themselves they would return to nothingness. Aquinas distinguished between a temporal beginning of the universe and its creation, where the latter concept refers to the existence of the universe as such. From this point of view an eternal yet created universe was perfectly possible. Even if the universe had no temporal beginning, it would still depend upon God's power for its very being.[13] Creation, Aquinas argued in *De Aeternitatis Mundi* from about 1270, had a double meaning:

> The first is that it presupposes nothing in the thing which is said to be created. … The second thing is that non-being is prior to being in the thing which is said to be created. This is not a priority in time or of duration, such as that what did not exist before does exist later, but a priority of nature, so that, if the created thing is left to itself, it would not exist, because it only has a being from the causality of the higher cause.

---

[12] For Philoponus's arguments, see Richard Sorabji, *Time, Creation and the Continuum: Theories in Antiquity and the Early Middle Ages* (Chicago: University of Chicago Press, 1983). Modernized versions of the infinity paradoxes of the Greek thinker have continued to attract attention. They were used in the 1970s as an argument against the steady-state theory of the universe. See Gerald Whitrow, "On the impossibility of an infinite past," *British Journal for the Philosophy of Science* **29** (1978), 39-45.

[13] See William Carroll, "Thomas Aquinas and big bang cosmology," *Sapientia* **53** (1998), 73-95, from which the quotation is taken. See also Conrad Hyers, *The Meaning of Creation: Genesis and Modern Science* (Atlanta: John Knox Press, 1984), and Copan and Craig, *Creation Out of Nothing*, pp. 147-157 (ref. 5).

Bonaventure, a contemporary Franciscan theologian, argued against Aquinas that the eternity of the world was heretical as well as philosophically absurd; because, had the world existed in an eternity the number of revolutions in the heavens must have been infinite, and for this reason the present could never have been reached. But it was Aquinas and not Bonaventure who won the discussion. Continual but timeless creation was eventually adopted by the Catholic Church under the name *creatio continua* (as opposed to *creatio originans*).

### 3. From Kant to Einstein

During the scientific revolution, roughly the 150-year period between Copernicus and Newton, astronomy was established as a branch of mechanical physics. Telescopic observations greatly expanded the astronomers' horizon, but the progress was basically limited to the solar system. Generally, cosmology played very little role and cosmogony even less. The question of the origin of the world was largely a non-question in the sense that everyone agreed that of course the world was divinely created. Characteristically, when the great astronomers of the period addressed the issue of the time of creation – and many of them did – they looked to biblical chronology rather than trying to answer the question by scientific means. Johannes Kepler found in this way that God had created the universe 3983 BC; the Danish astronomer Longomontanus, a pupil of Tycho Brahe, arrived at 3967 BC.

Whereas the date of creation could only be inferred from the Bible, a finite-age universe could be argued without it. If the cosmos were a machine slowly running down, such as Newton came to believe, it could not have existed forever, for in that case it would already be in a state of total dissolution (which it is not). As the British astronomer James Ferguson expressed it in a book of 1757, "For, had it existed from eternity, and been left by the Deity to be governed by the combined actions of the above [Newtonian] forces or powers, generally called Laws, it had been at an end long ago."[14] It was neither the first nor the last time that a counterfactual argument of this kind was used as evidence for a universe of finite age. As mentioned, it can be found much earlier in Lucretius' *De Rerum Natura* and other of the sources of ancient natural philosophy.

---

[14] Quoted in Helge Kragh, *Entropic Creation: Religious Contexts of Thermodynamics and Cosmology* (Aldershot: Ashgate Publishing, 2008), p. 19, where other examples of the argument can be found.

The traditional view was that God had created the universe more or less as it still is, but during the era of the Enlightenment this view was challenged by a series of evolutionary cosmogonies. According to these scenarios, the world as presently observed was the outcome of a slow evolutionary process starting in a very different state, perhaps a primordial chaos of the kind that the ancient atomists had assumed. The most innovative and elaborated version of the evolutionary cosmogonies was published in 1755 by 31-year-old Immanuel Kant as *Allgemeine Naturgeschichte und Theorie des Himmels* (Universal History and Theory of the Heavens). Kant started with a primeval, divinely created chaos of particles and then, ostensibly relying on the principles of Newtonian mechanics, explained how the chaotic state naturally evolved into condensations out of which the solar system and indeed the whole ordered universe was formed. What matters in the present context is that Kant's cosmic creation was an evolutionary process allegedly governed by the laws of physics and thus quite different from creation once and for all. "Creation is not the work of a moment," he emphasized. "Creation is never completed. Though it has once started, it will never cease. It is always busy in bringing forth more scenes of nature, new things and new worlds."[15]

While young Kant optimistically believed that the universe as a whole would be subject to scientific analysis, apparently he changed his mind. In his far better known and more influential *Kritik der reinen Vernunft* (Critique of Pure Reason) of 1781 he concluded that the notions of age and extent were meaningless when applied to the universe. In his so-called first antimony he first proved by means of logical arguments that "The world has a beginning in time, and is limited also with regard to space;" he next proved the opposite, that is, "The world has no beginning and no limits in space, but is infinite, in respect both to time and space."[16] Since the concept of the world at large was thus contradictory, he concluded that it cannot cover a physical reality but only be a concept of heuristic value. It was what he called a regulative principle. Kant's cosmogony of 1755 became the backbone of the later nebular hypothesis, also known as the Kant-Laplace hypothesis, which played a very important role during the nineteenth century. The role was controversial as well, for the hypothesis of a nebular origin of the universe, with no explanation of the original nebular stuff, was sometimes taken to imply an eternal and uncreated universe.[17] Evolution did not square easily with creation.

---

[15] Immanuel Kant, *Universal Natural History and Theory of the Heavens*, translated and edited by Stanley L. Jaki (Edinburgh: Scottish Academic Press, 1981), p. 155.

[16] *Critique of Pure Reason*, Chapter II, Section II.

In 1858 the German astronomer Johann Mädler suggested an argument for the origin of the universe a finite time ago which relied on observation rather than logic. His aim was not so much to provide evidence for a created universe, which he took for granted, as it was to explain the so-called Olbers' paradox of the darkness of the sky at night. The name refers to another German astronomer, Heinrich Wilhelm Olbers. As had been pointed out as early as the seventeenth century, if the universe was infinitely (or just enormously) large and filled uniformly with stars, the accumulated starlight should make the sky at night as bright as on a sunny day. And yet the night is dark. Mädler's solution was to combine the finite velocity of light with the hypothesis that the stars had not always existed. "If we knew the moment of creation, we should be able to calculate its boundary," he wrote, referring to the stars most far away.[18] However, his suggestion attracted almost no attention and was only revived much later in connection with the expanding universe discovered around 1930.

As mentioned, the question of the origin of the universe was given little priority by the astronomers. But it was considered interesting by the philosophers who in the spirit of Kant analysed it from a logical and conceptual point of view. One of them was Herbert Spencer, a prominent evolutionary philosopher in favour of the nebular world view. Spencer distinguished between three assumptions concerning the origin of the universe in an absolute sense. We may assert, he wrote in *First Principles* first published in 1862, that the universe is self-existent, or that it is self-created, or that it is created by an external agency. His analysis of the three possibilities led him to a conclusion no less pessimistic than the one Kant had arrived at more than eighty years earlier:

---

[17]  Stephen G. Brush, "The nebular hypothesis and the evolutionary worldview," *History of Science* **25** (1987), 245-278; Ronald L. Numbers, *Creation by Natural Law: Laplace's Nebular Hypothesis in American Thought* (Seattle: University of Washington Press, 1977).

[18]  Johann Mädler, *Der Fixsternhimmel* (Leipzig: Brockhaus, 1858), as quoted in Frank J. Tipler, "Johann Mädler's resolution of Olbers' 'paradox´," *Quarterly Journal of the Royal Astronomical Society* **29** (1988), 313-325. For the complex history of Olbers' paradox, see Stanley L. Jaki, *The Paradox of Olbers' Paradox* (New York: Herder and Herder, 1969) and Edward Harrison, *Darkness at Night: A Riddle of the Universe* (Cambridge, MA: Harvard University Press, 1987).

> Thus these three different suppositions respecting the origin of things, verbally intelligible though they are, and severally seeming to their respective adherents quite rational, turn out, when critically examined, to be literally unthinkable. It is not a question of probability, or credibility, but of conceivability. … Impossible as it is to think of the actual universe as self-existing, we do but multiply impossibilities of thought by every attempt we make to explain its existence.[19]

Cosmic creation might be beyond human comprehension, but it did not prevent scientifically based arguments for the universe being of finite age. The discovery in the 1850s of the second law of thermodynamics implied a tendency of all natural processes towards an equilibrium state of uniform temperature. If extrapolated to the far future, it indicated that all activity in the universe would come to an end, a state known as the "heat death." If extrapolated to the far past, it indicated that the universe had a beginning in time – or rather that there was a beginning for the operation of the laws of nature. The argument was often stated in terms of Rudolf Clausius' concept of entropy, a quantity which is a measure of degradation and has the unique property that it always increases in a closed system. The "entropic creation argument" can be stated counterfactually: if the universe had existed in an eternity of time, the entropy must now have reached its maximum; but since there is order and structure in the universe, this is obviously not the case; it follows that the age of the universe is finite, meaning that it had a beginning.[20]

This kind of argument, often supplied with the apologetic assumption that cosmic beginning implied divine creation, was hotly debated from about 1865 to 1915, but more among philosophers and theologians than among astronomers. It did not succeed in making a universe of finite age generally accepted. As late as 1913 the eminent British geophysicist Arthur Holmes referred to the entropic creation argument as follows:

> If the development of the universe be everywhere towards equalization of temperature implied by the laws of thermodynamics, the question arises – why in the abundance of time past, has this melancholy state not already overtaken us? Either we must believe in a definite beginning, in the creation of a universe furiously ablaze with energy, or

---

[19]  Herbert Spencer, *First Principles* (New York: P. F. Collier & Son, 1902), pp. 49-50.
[20]  See Kragh, *Entropic Creation* (ref. 14) for a full account of the argument and its history.

else we must assume that the phenomena we have studied simply reflect our limited experience.[21]

However, Holmes denied the validity of the argument and maintained that the universe as a whole had existed eternally and probably evolved through an infinite number of cycles.

At the time of World War I cosmology did not yet exist as a scientific discipline. In so far that astronomers dealt with cosmological questions they focused on the structure of the stellar universe, in particular the size of the Milky Way and its relation to the nebulae. Were the nebulae parts of the Milky Way system or were they huge conglomerates of stars, milky ways in their own right, far away from it?

In 1917 Albert Einstein laid the foundation of modern cosmology by proposing a model of the universe on the basis of his new theory of gravitation, the general theory of relativity.[22] Although Einstein's model was a revolution in cosmological thought, its picture of the universe was in some sense traditional. The model presupposed that the universe as a whole was uniform and spatially closed corresponding to a positive curvature of space; it was finite yet with no boundary and therefore contained but a finite number of stars. Importantly, it was also static in the sense that the curvature of space and the mean density of matter remained constant. To maintain a static universe in accordance with astronomical observations Einstein had to introduce a new term in his cosmological field equations, the later so famous cosmological constant. Being static his universe had no temporal dimension but was eternal in both past and future time. For this reason alone the question of the origin of the universe did not enter Einstein's mind. Nor did it enter the minds of the few other physicists and astronomers occupying themselves with his mathematically and conceptually abstruse theory.

### 4. The big bang hypothesis

It turned out that Einstein's cosmological field equations were much richer in mathematical structure than he thought at first. The equations do not merely describe a static universe of the type Einstein examined in 1917 but also a whole class of dynamical models, that is possible universes with a curvature (and hence size) that varies in time. The first to point out these mathematical solutions was the Russian physicist Alexander Friedmann, who in a paper of 1922 analysed all uniform models described by the field equations. Some of these models were expanding, meaning that the size of the universe (as given by the distance measure $R$) increased in cosmic time $t$. As Friedmann realized, there were models with the remarkable property that $R = 0$ for $t = 0$ in the past, or what he called a "beginning of the world." $R = 0$ corresponds to a point in space, a "singularity" with no spatial extension at all. Friedmann wrote: "The time since the creation of the world is the time which has passed from the moment at which space was a point ($R = 0$) to the present state ($R = R_0$)."[23]

Here we have, for the first time, the notion of the origin of the universe derived not from a philosophical doctrine but from a fundamental theory of physics. On the other hand, Friedmann's brilliant investigation was primarily a mathematical exercise and he did not express any preference for one model over another. He did not argue that our universe is *in fact* expanding or that it *really* had an origin in a singularity. At any rate, for nearly a decade his paper remained either unknown or unappreciated.

By 1930 the expansion of the universe, now supported by observation as well as theory, had become a reality. But the notion of a beginning of the world does not follow logically from cosmic expansion. What became known as the big bang universe in a realistic sense was first proposed on 9 May 1931 in a brief note in the journal *Nature*. The author was Georges Lemaître, a 36-year-old Belgian astrophysicist and cosmologist who was also trained as a Catholic priest.

"We could conceive," Lemaître wrote in his 1931 paper, "the beginning of the universe in the form of a unique atom, the atomic weight of which is the total mass of the universe … [and which] would divide in smaller and smaller atoms by a kind of super-radioactive process."[24] He used the term "atom" in a metaphorical sense close to that of the ancient Greeks, namely as something completely undifferentiated and devoid of physical properties. Moreover, he carefully spoke about the beginning or origin of the universe, not its creation. As a faithful Catholic Lemaître was convinced that God had *created* the universe, and yet he stressed that its *origin* was a natural event and that his theory was purely scientific. At one occasion he wrote that "the hypothesis of the primeval atom is the anti-thesis of the supernatural creation of the world." He elaborated:

> We may speak of this event as of a beginning. I do not say a creation. Physically it is a beginning in the sense that if something has happened before, it has no observable influence on the behaviour of our universe. … Any pre-existence of our universe has a metaphysical character. Physically everything happens as if it was really a beginning. The question if it was really a beginning or rather a creation, something starting from nothing, is a philosophical question which cannot be settled by physical or astronomical considerations.[25]

---

[23] Alexander Friedmann, "Über die Krümmung des Raumes," *Zeitschrift für Physik* **10** (1922), 377-386; Harry Nussbaumer and Lydia Bieri, *Discovering the Expanding Universe* (Cambridge: Cambridge University Press, 2009).

[24] G. Lemaître, "The beginning of the world from the point of view of quantum theory," *Nature* **127** (1931), 706. For details and perspectives, see Helge Kragh and Dominique Lambert, "The context of discovery: Lemaître and the origin of the primeval-atom hypothesis," *Annals of Science* **64** (2007), 445-470.

According to Lemaître's scenario, at $t = 0$ the universe already existed in the shape of what he called a "primeval atom," a relatively small body of enormous mass density. Such a hypothetical super-atom was comprehensible if not subject to scientific analysis. Lemaître insisted that it was physically meaningless to speak of time before the initial explosion and yet he wrote, inconsistently it seems, "the beginning of the world happened a little before the beginning of space and time." But if time came into being only with the original explosion (the big bang), how could the world have begun "a little before"? According to Lemaître, immediately after the disintegration of the primeval atom it would, at least in principle, be possible to analyse the very early universe by means of the laws of physics. Whereas he considered the primeval atom to be real, he denied that the cosmic singularity $R = 0$ formally turning up in the equations at $t = 0$ could be ascribed physical reality. The "annihilation of space," as he called it, was for him a mathematical artefact.

Well acquainted with the philosophical classics Lemaître knew about Kant's argument against a cosmic beginning. If the universe started with the explosion of the primeval atom, what caused the explosion? Cause precedes the effect, so how can there be a causal agent *before* the beginning when time did not even exist? Lemaître admitted that Kant's objection was a genuine dilemma in so far that the universe is governed by the principles of causality and determinism inherent in classical mechanics. But in quantum mechanics processes can occur without a cause, such as is the case in radioactive decay. Lemaître considered the quantum origin of the universe to be essential, for only in this way could one avoid the tricky question of what caused the initial disintegration. Although his idea was received with interest in the popular press, astronomers either ignored or rejected it. A Canadian astronomer characterized it as "the wildest speculation of all" and "an example of speculation run mad."[26] And yet it was out of this wild speculation that the modern theory of the big bang universe emerged.

---

[25] Quoted in H. Kragh, *Matter and Spirit in the Universe: Scientific and Religious Preludes to Modern Cosmology* (London: Imperial College Press, 2004), pp. 147-148.

In the late 1940s Lemaître's daring hypothesis was independently transformed into a more detailed and advanced theory of the early universe, now by conceiving it in terms of nuclear physics. According to the Russian-American physicist George Gamow the very early universe was a hot and dense inferno of interacting nuclear particles, and as a result of the interactions the chemical elements were formed during a brief period in the cosmic past. Together with his collaborators Ralph Alpher and Robert Herman he realized in 1948 that the earliest inferno must consist predominantly of high-energy radiation rather than particles such as protons and neutrons. On this basis they succeeded in establishing the essential features of the "hot big bang" theory as it is known today. With respect to the absolute origin of the universe, Gamow intentionally disregarded it. He and his collaborators simply started their calculations in a pre-existing original universe, without concerning themselves with where it came from (except that Gamow speculated that the big bang might be the result of the collapse of a previous universe – a "big crunch"). Although his cosmological model was often labelled a creation theory, in reality it was an evolution theory. Gamow did use the term "creation," but merely in the innocent sense of "making something shapely out of shapelessness."[27]

Finite-age models of the type proposed by Lemaître and Gamow were challenged by the fundamentally different steady state theory of the universe introduced by Fred Hoyle, Hermann Bondi and Thomas Gold in 1948. According to this theory the universe had existed in an eternity of time and would continue existing eternally. Moreover, its average density of matter remained the same despite its continual expansion, which was explained by postulating a tiny amount of matter creation throughout the universe.[28] This element of spontaneous matter creation aroused heated debate. While some scientists objected that the fundamental law of energy conservation could not be violated, philosophers tended to conceive matter creation as an example of *deus ex machina* reasoning. To Mario Bunge, a physicist and philosopher, the steady state theory was nothing but "science-fiction cosmology."[29] Of course, Hoyle and his supporters denied the charges.

---

[26] John S. Plaskett, "The expansion of the universe," *Journal of the Royal Astronomical Society of Canada* **27** (1933), 235-252.

[27] G. Gamow, *The Creation of the Universe* (New York: Viking Press, 1952), preface.

What matters is that by assuming an infinite age of the universe the steady state theorists avoided the thorny question of a beginning. It was in this context that Hoyle, on 28 March 1949, gave a BBC broadcast in which he coined the name "big bang" for the kind of cosmological theory which assumed an origin of the universe in an explosive event. The following year he characterized "the big bang assumption [as] an irrational process that cannot be described in scientific terms."[30] What he had in mind was the old objection that there can be no causal explanation, indeed no explanation of any kind, for the beginning of the universe. At more than one occasion he associated the big bang theory with theism, suggesting that a temporal beginning of the universe implied divine creation and was therefore unscientific. For example: "The passionate frenzy with which the big-bang cosmology is clutched to the corporate scientific bosom evidently arises from a deep-rooted attachment to the first page of Genesis, religious fundamentalism at its strongest."[31]

During the period from 1948 to the early 1960s the steady state theory was a serious alternative to evolutionary models based on Einstein's equations, whether these were finite-age models or not. However, with the discovery of the cosmic microwave background in 1965 the balance tipped decisively to the advantage of the big bang. According to this theory, in the very early and hot universe light (or photons) would be coupled to elementary particles and unable to escape them. But when the universe cooled enough for protons and electrons to form hydrogen atoms the universe became transparent – filled with freely moving photons. This "background radiation" originating more than 13 billion years ago still exists in the form of very weak microwaves. The existence of a background radiation thus follows naturally from the big bang theory and had in fact been predicted on this basis by Alpher and Herman as early as 1948. On the other hand, the new phenomenon could be accommodated by the steady state theory only by means of arbitrary and highly artificial hypotheses.

---

[28] See H. Kragh, *Cosmology and Controversy: The Historical Development of Two Theories of the Universe* (Princeton: Princeton University Press, 1996).

[29] M. Bunge, "Cosmology and magic," *The Monist* **47** (1962), 116-141.

[30] Fred Hoyle, *The Nature of the Universe* (New York: Harper & Brothers, 1950), p. 124. On the origin and history of the term "big bang," see H. Kragh, "Naming the big bang," *Historical Studies in the Natural Sciences* **44** (2014), 3-36.

[31] F. Hoyle, "The universe: Past and present reflections," *Annual Review of Astronomy and Astrophysics* **20** (1982), 1-35, p. 23.

To make a long story short, by the late 1960s the steady state theory was practically dead and the big bang alternative accepted by the majority of physicists and astronomers. Today a much-refined version of Gamow's hot big bang cosmology has the status of a paradigm. As regards the name "big bang" some leading cosmologists have suggested that it is a misnomer. This is not only because it alludes to a noisy explosion localized in space but also because the big bang, if taken to be a creation event at $t = 0$, is outside the standard models of physics and cosmology.[32]

Hoyle remained throughout his life a sharp critic of the victorious big bang theory. Not only did he find it methodologically objectionable, he also argued that an explosive beginning in a very simple object failed to account for the evolution of order and structure in the universe. Here is how he phrased his objection in an address of 1993, realizing that his opposition to the big bang theory was shared by only a small minority of his colleagues:

> Explosions do not usually lead to a well-ordered situation. An explosion in a junk-yard does not lead to sundry bits of metal being assembled into useful working machines. Yet after expanding for about a billion years something of this nature is supposed to have happened to the universe. Galaxies formed that are widely similar over large volumes of space. Stars formed. Life originated and evolved. Man arose and began to think about it all. How such a structured world came into being remains unexplained.[33]

According to modern big bang cosmology there is no basis for Hoyle's objection. On the contrary, structures in the universe follow from the lack of homogeneity that the theory predicts for its very early development.

## 5. Aspects of modern cosmology

The big bang standard theory has since its establishment some fifty years ago been greatly developed theoretically as well as observationally. The best cosmological

---

measurements combined with the best theoretical model results today in an age of the universe of remarkable accuracy, namely 13.799 ± 0.021 billion years. Some of the advances relate to the very early universe and even to the ultimate question of its origin. Did the universe originate in an extended object of some kind, perhaps a modern analogue of Lemaître's primeval atom, or in a singularity of zero extension? If the latter is the case it is tempting to identify the initial singularity with the absolute beginning of the universe. Since the singularity has no physical properties whatsoever, the question of where it came from need not arise.

In work around 1965 Roger Penrose, Stephen Hawking and a few other mathematical physicists proved that a universe governed by the general theory of relativity must necessarily possess a space-time singularity. Although the proof is sometimes taken to imply that the universe started in a singularity, this is too strong an interpretation. Almost all cosmological models describe a finitely old universe but not, when physics is added to the mathematics, a universe starting in a singularity. The Penrose-Hawking singularity theorem builds solely on general relativity and thus does not take quantum effects into regard. But it is generally believed that the physics of the very early universe can be understood only on the basis of a unified theory of gravitation and quantum mechanics. This era of quantum gravity is believed to be the inconceivably small time interval between $t = 0$ and $t = 10^{-43}$ sec, where the latter is known as the Planck time. As far as physics is concerned, the initial singularity is not inevitable.

In modern early-universe cosmology the vacuum is a main player. From the point of view of quantum mechanics the vacuum is entirely different from the void or nothingness of classical physics (or perhaps one should say metaphysics). A vacuum is necessarily filled with energy which fluctuates wildly and spontaneously. Might the universe have its origin in a quantum fluctuation? This is what the American physicist Edward Tryon proposed in 1973, thus giving a new twist to the concept of *creatio ex nihilo*. Although Tryon's model turned out to be flawed, other physicists came up with similar suggestions of "Creation of Universes from Nothing," as the title of a 1982 paper reads.[34] The author, the Russian-American physicist Alexander Vilenkin, claimed that his theory explained how "the universe is spontaneously created from literally *nothing*." However, this and other theories in the same tradition do not really explain the creation of the universe *ex nihilo* since they merely push back the creation scenario to a hypothetical vacuum scenario. They presuppose a primordial quantum vacuum, which is an object describable by the laws of physics and in no way the same as "nothing." Where did the quantum vacuum come from?

---

[34] E. Tryon, "Is the universe a vacuum fluctuation?" *Nature* **246** (1973), 396-397; Alexander Vilenkin, "Creation of universes from nothing," *Physics Letters* **117 B** (1982), 25-28. See also Alexei Starobinsky, "Future and origin of our universe: Modern view," pp. 71-84 in V. Burdyusha and G. Khozin, eds., *The Future of the Universe and the Future of Our Civilization* (Singapore: World Scientific, 2000).

At the time Vilenkin wrote his paper the picture of the very early universe had undergone a revolution in the form of the so-called inflation theory. To put it briefly, according to this theory there was an extremely brief phase in the history of the *very* early universe, shortly after the Planck time, in which empty space expanded at a stupendous speed. Although the inflation lasted from only $10^{-36}$ sec to about $10^{-33}$ sec after $t = 0$, during this brief interval of time space expanded by a factor of $10^{30}$ or more. The basic mechanism responsible for the huge expansion is believed to be a hypothetical "inflaton field" which can be represented by a quantum version of the cosmological constant appearing in Einstein's equations. This constant has the remarkable property that it leads to a negative pressure and an associated vacuum energy *density* (rather than the energy itself) which remains constant. It follows that the inflation generates an enormous amount of energy – almost out of nothing. After the brief inflationary phase, the much slower normal expansion of the now very hot and energy-rich space takes over.

It all sounds very exotic, almost incredible, but most cosmologists consider the inflation scenario, in one of its many versions, to be convincing because of its explanatory and predictive power. They believe that we know what the universe looked like just $10^{35}$ sec after $t = 0$. This is most interesting but cynics will argue that it does not bring us nearer to answering the question of the ultimate origin. Where did the inflaton field come from?

The success of the inflation theory drew increased attention to the role of the cosmological constant as a measure of the energy density of the vacuum. However, it was not studies of the very early universe that confirmed Einstein's cosmological constant but astronomical observations of the present expansion rate. In the late 1990s it turned out that the universe is accelerating, meaning that it expands at an increasing rate as if it is blown up by a self-repulsive "dark energy." The precise nature of this dark energy is still unknown but the consensus view is that it is a manifestation of the vacuum energy associated with the cosmological constant. This strange kind of energy actually dominates the present universe, as it makes up roughly two-thirds of all energy and matter in the universe and in the future will dominate even more. While the discovery of dark energy has great consequences for the far future of the universe it is not equally relevant for the very early universe. On the other hand, it underlines the importance of the vacuum energy density as a fundamental characteristic of our universe.

The original inflation theory was soon developed into versions of "eternal inflation" primarily by Vilenkin and Andrei Linde, who suggested that in the universe as a whole, new inflating regions will be produced more rapidly than non-inflating regions. Inflation is self-generating, if not in our observed universe then in the much bigger and presumably infinite universe at large. According to Linde, "the universe is an eternally existing, self-reproducing entity that is divided into many mini-universes much larger than our observable portion, and … the laws of low-energy physics and even the dimensionality of space-time may be different in each of these mini-universes."[35]

Here we have the controversial and much-discussed hypothesis of the so-called *multiverse*, the idea that there exists a multitude of other universes each with its own vacuum energy density.[36] We also have the no less controversial idea that, despite the big bang origin of our universe, the universe at large is infinite in its temporal extension. According to some proponents of eternal inflation the infinity covers the past as well as the future, meaning that there is no proper origin. Other proponents argue that although inflation will go on forever in the future, it is probably not eternal in the past. In that case a primary big bang is still part of the picture.

As several modern cosmologists have noted, the classical and by now defunct steady state theory of Hoyle and his colleagues can in some respects be seen as a precursor of eternal inflation, but only if the latter theory can be extended eternally to the past. For those who think that it can, Hoyle's motivation is still valid. In a paper advocating eternal inflation in the past two physicists say about the steady state theory that it is "appealing because it avoids an initial singularity [and] has no beginning in time."[37]

---

[35]  A. Linde, *Inflation and Quantum Cosmology* (Boston: Academic Press, 1990), p. 29.

[36]  The multiverse hypothesis exists in several versions. According to some of them the many other universes are causally separate, meaning that they are unobservable even in principle. For an account of the multiverse and its history ca. 1990-2010, see H. Kragh, *Higher Speculations: Grand Theories and Failed revolutions in Physics and Cosmology* (Oxford: Oxford University Press, 2011), pp. 291-324.

## 6. Did the universe have a beginning?

Eternal inflation is one way of avoiding what some physicists consider unpalatable, namely an absolute origin of the universe, but there are several other ways. One of them is to introduce a new measure of time that conjures an infinite past out of the finite one. This can be done by replacing the ordinary time parameter $t$ by a new one $\theta$ which is logarithmically related to $t$ time. On the new concept of time the big bang did not occur at $t = 0$ but at $\theta = -\infty$ (minus infinity), which means that it never *began*. In the words of a French physicist, "the numerical finiteness of the age of the universe by no means precludes its conceptual infiniteness." There is a sense in which "the universe is infinitely old and had no definite beginning."[38] In this sense the big bang can be approached asymptotically but never reached, somewhat in analogy to the concept of zero absolute temperature (see Section 7). Conceptually appealing as the idea may seem, most physicists consider it nothing but a formal trick. They maintain that there was an original big bang approximately 14 billion years ago.

In philosophical and mythological contexts the idea of a cyclic or oscillating universe can be found in ancient Greek and Indian cosmologies. The general idea is that our present universe is the outcome of a previous one and that it will itself result in a successor universe; and, moreover, that there is an endless number of these earlier and later universes. If so there would be neither a beginning nor an end to the universe as a whole. In more or less scientific versions ever-cyclic models attracted much interest during the nineteenth century, sometimes in the form of the eternally recurrent universe in which the cycles were identical in every detail and every single event in history thus endlessly repeats itself. For example, the famous German philosopher Friedrich Nietzsche advocated such a world view.[39]

---

[37]  Anthony Aguirre and Steven Gratton, "Steady-state eternal inflation," *Physical Review D* **65** (2002), 083507.

[38]  Jean-Marc Lévy-Leblond, "Did the big bang begin?" *American Journal of Physics* **58** (1989), 156-159. The idea of two time scales goes back to the 1930s, when it was first discussed by the British cosmologist Edward Arthur Milne, who distinguished between what he called kinematic and dynamic time.

[39]  Robin Small, *Nietzsche in Context* (Aldershot: Ashgate, 2001).

In the context of relativistic cosmology cyclic models were reconsidered by Friedmann in his pioneering paper of 1922. In this context the expansion of the universe is followed by a contracting phase which again is followed by an expanding phase, and so on. The bounce from contraction to expansion – or from a big crunch to a big bang – is supposed to take place smoothly between two non-singular states of high but not infinitely high density and temperature. Many models of this kind have been proposed but they all suffer from severe difficulties and especially so if they suppose an infinite number of earlier universes[40]. If there can only be a finite number of previous cycles the whole idea loses much of its philosophical appeal, for then there must be a first cycle and the questions of its origin reappears. The classical cyclic universe only makes sense if space is closed and expanding at a decreasing rate. With the discovery of the acceleration of space this turned out to be wrong and models of this kind were consequently abandoned.

Nonetheless, the appeal of an endless universe with no beginning was too strong to be given up completely. In a new cyclic model proposed in 2002, Paul Steinhardt and Neil Turok developed a cosmology with an eternal sequence of identical cycles consisting of expansions and contractions. Contrary to earlier models it relied on an open and accelerating universe in agreement with observations and it was specifically constructed as an alternative to the inflation scenario. In a popular address Steinhardt summarized: "Space and time exist forever. The big bang is not the origin of time. Rather, it is a bridge to a pre-existing contracting era. The universe undergoes an endless sequence of cycles in which it contracts in a big crunch and re-emerges in an expanding big bang, with trillions of years of evolution in between."[41] Although a few physicists continue developing the model, it has failed in making an impact on mainstream cosmology. Yet it is worth mentioning as a modern example of the enduring appeal of an eternal-cyclic universe in which the question of origin does not arise.

Finally there is a class of cosmological models which include the hypothesis of a pre-big bang universe but are not cyclic in the ordinary sense. One may speak of bouncing rather than cyclic models. At about 1950 Gamow speculated that the universe might have evolved from a previous state of nearly infinite rarefaction which slowly had contracted gravitationally into a super-dense state; out of this state the big bang of our present universe emerged. It was thus a temporally symmetric picture of the universe, stretching from minus infinity to plus infinity and with no absolute beginning. Modern theories founded on quantum gravity have resulted in pictures which, from a qualitative point of view, share the basic features of Gamow's classical picture.

---

[40] For the history of cyclic universe models, see Kragh, *Higher Speculations* (ref. 36), pp. 193-216.

[41] P. Steinhardt, "The endless universe: A brief introduction," *Proceedings of the American Philosophical Society* **148** (2004), 464-470. See also P. Steinhardt and N. Turok, *Endless Universe: Beyond the Big Bang* (New York: Doubleday, 2007).

The best offer of a theory of quantum gravity may be the fundamental theory of superstrings, a unified many-dimensional theory of gravity and the three forces of nature which can be understood in terms of quantum mechanics. These are the well-known electromagnetic force and the two short-range forces known as the weak and the strong (or nuclear) interactions. It turns out that the electromagnetic and the weak forces can be unified in a single theory and that this "electroweak" theory can be extended to cover also the strong force in what is called "grand unified theory." String theory is even grander as it offers a unified formalism encompassing all the four interactions. The prefix "super" indicates so-called supersymmetry with the implication that all the known elementary particles having partner particles. For example, the photon has a supersymmetric partner called a photino. Alas, none of these superpartners have been detected in experiments. What is of relevance here is that physicists have constructed cosmological models on the basis of string theory and that these models avoid the initial singularity and the problem of an absolute beginning.

According to string cosmology or what is sometimes called pre-big bang cosmology, the big bang at $t = 0$ was not the origin of everything but a moment in cosmic time when a state of very high but finite density bounced into a state of rapidly decreasing density. Strings are irreducible one-dimensional objects and the theory includes a fundamental length which can be thought of as the dimension of a point in space; the length is about $10^{-34}$ m, which is also the smallest radius of cosmic space. The string scenario posits a flat and nearly empty universe in the indefinite past and also, symmetrically, in the indefinite future. From the eternally existing pre-universe our universe emerged when the density reached the maximum value in a big crunch, or what from our point of view was a big bang. After that followed an inflationary phase and eventually the eternally accelerating universe as we observe it today.

Although there is no definite origin of the string universe, one of its leading advocates, the Italian physicist Maurizio Gasperini, believes that "the Universe was born according to God's will, with an act of creation having its ultimate and complete purpose in human beings."[42] He recognizes that this is of course a personal view and not one which can be justified scientifically. There was no origin of the universe, and yet it was created. His view brings to mind the old discussion going back to Thomas Aquinas of whether or not God could have created an eternal universe.

Apart from string theory, "loop quantum gravity" (LQG) is another candidate for a unified theory of general relativity and quantum mechanics. LQG is entirely different from string theory, for other reasons because it does not include supersymmetry and operates with only the four known dimensions of space-time. Despite the differences, when applied to the universe at large it results in a picture which is surprisingly similar to the one of string cosmology.[43] While strings cannot be squeezed to zero volume, in LQG space itself is discrete, in a sense made up of minimum "space atoms" of a volume of the order of $10^{-100}$ m³. For this reason loop quantum cosmology reproduces the feature of string cosmology, that there is no big bang singularity. The universe pictured by LQG theorists also has no beginning and no end. It develops from a past-eternal pre-universe over the bounce at $t = 0$ to the future-eternal present universe. In spite of the similarities between the cosmic scenarios offered by the two theories of quantum gravity they result in different predictions which can in principle (but perhaps only in principle) be tested by measurements. Work in this or other  traditions of quantum cosmology continues to this day, suggesting that it is theoretically possible to explain the big bang at $t = 0$ as the result of a previously contracting universe.[44]

It seems that modern theories of quantum gravity are no more able than other cosmological theories to come up with a good answer concerning the ultimate origin of the universe. Here is the view of Thanu Padmanabhan, an Indian cosmologist and specialist in quantum gravity who has also investigated modern steady state theories of the universe:

---

[42]  M. Gasperini, *The Universe Before the Big Bang: Cosmology and String Theory* (Berlin: Springer, 2008), p. 195.

[43]  Martin Bojowald, "Follow the bouncing universe," *Scientific American* **299** (April 2008), 44-51.

[44]  Steffen Gielen and Neil Turok, "Perfect quantum cosmological bounce," *Physical Review Letters* **117** (2016), 021301.

> How (and why!) was the universe created and what happened before the big bang?
> The cosmologist giving the public lecture usually mumbles something about requiring
> a quantum gravity model to circumvent the classical singularity – but we really have
> no idea! String theory offers no insight; the implications of loop quantum gravity for
> quantum cosmology has attracted fair amount of attention recently but it is fair to say
> that we still do not know how (and why) the universe came into being.[45]

Proponents of pre-big bang cosmologies may respond that the universe never came
into being and thus there is no question to be answered.

## 7. Philosophical and theological perspectives

We live in a big bang universe, but unfortunately there is no consensus what the key
term "big bang" covers. Physicists and astronomers generally refer to the big bang as
a brief but crucial chapter in the history of the universe, say from $t = 10^{-12}$ seconds (or
sometimes $10^{-35}$ seconds) to $t = 10^4$ seconds. In this sense the big bang can be
considered a scientific fact supported by a wealth of reliable evidence. But the term is
also used, especially but not only by philosophers, as a reference to the absolute
beginning at $t = 0$. This is quite a different meaning and the concept it covers can in
no way be characterized as scientifically documented. Creation in an absolute and
therefore metaphysical sense is not part of what most astronomers and physicists
refer to as the big bang scenario any more than an absolute origin of life is part of
what most biologists refer to as the neo-Darwinian evolution scenario.

Terminological ambiguities apart, if there were an absolute origin at $t = 0$, a
cosmic creation event, can it be explained? The emergence of the first elementary
particles during the quark era some $10^{-9}$ seconds later can be explained from the
previous state of the universe, but what about the beginning of time itself? As
mentioned, an ordinary causal explanation based on an earlier state is out of the
question, at least if we disregard speculations of a multiverse or a pre-big bang
universe. As an alternative one might consider other forms of explanation not
ordinarily used in science, such as a teleological explanation where the origin of the
universe is associated with a purpose; or one may just renounce the possibility of
understanding the singular event. This event, the creation of the universe, concerns not only the origin of what constitutes the universe – space, time, energy, fields, and matter – but also the origin of the laws that govern it. Scientific explanations are nomological, meaning that they are based on laws of nature, but how did these laws come into being? When cosmologists describe the early universe they make use of relativity theory, quantum mechanics, thermodynamics and other theories. Why do these laws of nature apply rather than some other laws that one might imagine?

---

[45]  T. Padmanabhan, "Understanding our universe: Current status and open issues," pp. 175-
204 in A. Ashtekar, ed., *100 Years of Relativity. Space-Time Structure: Einstein and Beyond* (New
Jersey: World Scientific, 2005), p. 199.

It has been suggested that the singular big bang event is not something that belongs to the ontology of cosmic evolution but that it can be approached only asymptotically. Concepts of this kind are known from other areas of physics which may serve as analogies. For example, the absolute zero of temperature ($T = 0$ K or approximately $-273.15$ °C) can be approached arbitrarily closely but never be reached either in nature or in the laboratory. Like many analogies it serves a heuristic purpose only. A decisive difference is that the temperature $T = 0$ is well defined and can be understood in terms of physics. The cosmic $t = 0$ event, on the other hand, is not comprehensible in the same way. It is tempting to consider this event in the light of epistemology rather than ontology. In this case the singular big bang is not something which once existed but an epistemic horizon, a limit for the intelligibility of the cosmos. Science has always been faced with boundaries beyond which it seemed powerless but which were nonetheless transgressed by the progress of science. These boundaries or horizons move, and there is no reason why the Planck time is an absolute horizon of knowledge. Yet it is possible that at least one boundary, the one at $t = 0$, will never be removed.[46]

At the bottom of any discussion of the origin of the universe lies the difficult concept of time. Instead of imagining a metaphysical state of nothingness out of which the physical universe magically emerged, one may imagine a frozen proto-universe in which there were no processes at all and hence also no possibility of defining time. It may have been something of this kind that Lemaître had in mind when he introduced the primeval atom in 1931. However, the imagery seems to be of no help and is in any case beyond science. Time is relational. It presupposes an active universe with processes that can function as clocks and the clocks of course belong to the universe. According to Stephen Hawking, "To ask what happened before the universe began is like asking for a point on Earth at 91 north latitude; it just is not defined."[47] Literally speaking the big bang universe has *always* existed and will exist *forever* irrespective of a future big crunch or not. The terms "always" and "forever" are temporal terms meaning at all times and there was no time at which the universe did not exist.[48] The word "then" is also temporal and for this reason one cannot say that "first there was nothing and then there was something." Although the big bang universe has always existed it is not past eternal like the steady state universe.

---

[46] Willem B. Drees, *Creation: From Nothing until Now* (London: Routledge, 2002), pp. 13-15; Milton K. Munitz, *Cosmic Understanding: Philosophy and Science of the Universe* (Princeton: Princeton University Press, 1986), p. 172.

It should be kept in mind that time as used by cosmologists is basically a mathematical parameter appearing in the field equations. The time parameter $t$ is positive and varies continuously, meaning that it can attain any value however small. In principle it can be immensely smaller than the Planck time $10^{-43}$ seconds. Why not $10^{-1000}$ seconds? The way that specialists in early-universe cosmology speak about time is not only abstract but also in many ways remote from the usual concept where time is a measure associated with real physical processes. It is generally believed that the cosmic time scale must have a physical basis in the form of a clock and not be just a mathematical symbol. The problem is that there are no known processes varying with a period so small as the Planck time or the time scale of the inflation era. The smallest decay time known from particle physics is about $10^{-24}$ seconds, trillions of times larger than the Planck time. It is far from clear if the time concept used in this branch of cosmology is well defined or is the same concept of time as used elsewhere in science.[49] And it is even less clear if the physical concept of time can be meaningfully extended to the pre-big bang universe hypothesized by some theories of quantum gravity.

The question of the origin of the universe continues to be of theological interest and perhaps even more so than of scientific interest.[50] On the face of it the big bang theory with its cosmic origin of time seems to offer support for a theistic interpretation. After all, if the origin at $t = 0$ is beyond scientific comprehension and the universe is nonetheless of finite age, doesn't it confirm the view of traditional theism that God, a necessary being, created the universe out of nothing? Although the success of the big bang model does not really confirm theism, this is what many believers have argued. Notably, in an address of 1951 the pope, Pius XII, endorsed the new big bang theory as scientific proof of the biblical creation story.[51] "Everything seems to indicate that the material content of the universe had a mighty beginning in time," the pope asserted. He continued:

> Thus, with that concreteness which is characteristic of physical proofs, it [modern science] has confirmed the contingency of the universe and also the well-founded deduction as to the epoch when the world came forth from the hands of the Creator. Hence, creation took place. We say: therefore, there is a Creator. Therefore, God exists!

The papal argument presupposes cosmic creation *ex nihilo* to be miraculous and yet at the same time it derives creation from a scientific theory based on the laws of nature. The theist cannot have it both ways. God transcends the laws of nature, which he created together with the universe, and the laws cannot act as a guide to how he created the universe.

The question of the relationship between big bang cosmology and divine creation has been discussed endlessly but without much coming out of the discussion.[52] Just as the finite-age big bang universe offers no strong support for theism, so an infinitely old universe offers no strong support for atheism. The American astronomer Carl Sagan evidently thought that an eternal universe is incompatible with theism. Imagine that the pre-big bang model or some other past-eternal model was proved correct. This, Sagan suggested, would be "the one conceivable finding of science that could disprove a Creator – because an infinitely old universe would never have been created."[53] However, to counter the argument the theist only has to appeal to the concept of continual creation discussed in the middle ages. Irrespective of its age there needs to be a cause that maintains the existence of the universe and this cause qualifies as a creation.

---

[51] The papal address can be found in P. J. McLaughlin, *The Church and Modern Science* (New York: Philosophical Library, 1957), pp. 137-147, and online as http://www.papalencyclicals.net (accessed 10/08/2016).

[52] See, for example, William L. Craig and Quentin Smith, *Theism, Atheism and Big Bang Cosmology* (Oxford: Clarendon Press, 1995).

[53] C. Sagan, *The Demon-Haunted World: Science as a Candidate in the Dark* (London: Headline, 1997), p. 265.

Indeed, most theologians and Christian philosophers agree that *creatio continua* is more important than the original form of creation corresponding to a finite-age universe. As early as 1933 the British mathematician and bishop of Birmingham, Ernest William Barnes, pointed out that his God was not to be found in the origin of our universe:

> Men have thought to find God at the special creation of their own species, or active when mind or life first appeared on the earth. They have made him God of the gaps in human knowledge. To me the God of the trigger is as little satisfying as the God of the gaps. It is because throughout the physical Universe I find thought and plan and power that behind it I see God as creator.[54]

During the controversy over steady state cosmology in the 1950s this theory, with its infinite time scale in both directions, was sometimes considered a challenge to theism. But as theologians were quick to point out, the claimed problem was nothing but a pseudo-problem; the question of whether the universe has a beginning or not, is of no real importance for theology. A later theologian expressed it in this way: "Divine creativity is not restricted to a finite stretch of time, or to the past, but is a continuing activity, as theologians from Augustine to Luther and Calvin to the present have argued. Creation is not just a matter of beginnings."[55] It is debatable whether this kind of response is satisfactory. Continual creation in the theological sense is metaphysical and without a counterpart in physical cosmology. Sceptics will argue that the appeal to *creatio continua* is nothing but a way of protecting the Christian creation doctrine from scientific criticism. If divine creation is simply based on the undeniable existence of the world it is impossible to argue against the doctrine.

There is no agreement between theologians, or between theologians and cosmologists, regarding the relationship between Christian belief and cosmological models. Most theologians deny a direct relation, such that the finite-age big bang universe lends support to the hypothesis of a divine creator or that an eternal universe disproves the hypothesis. But not all do. The Christian philosopher William

---

[54] E. W. Barnes, *Scientific Theory and Religion* (Cambridge: Cambridge University Press, 1933), p. 409.

[55] Hyers, *The Meaning of Creation* (ref. 13), p. 67. For the response to the steady state theory, see Erich L. Mascall, *Christian Theology and Natural Science* (London: Longmans, Green and Co., 1956).

Craig sides with Pius XII when he claims that "the big bang model thus provided dramatic empirical verification of the biblical doctrine of *creatio ex nihilo*."[56]

The question has been discussed for more than thousand years and one might perhaps expect that the recent progress in physical cosmology would have led to some clarification or even definite answers. But this has not been the case and it is unlikely that it will ever happen. Science is impotent with respect to theological doctrines and theology is of no direct relevance to science. Let me end this essay by quoting the balanced view from an introductory textbook in astronomy:

> If we use God as an explanation for the big bang, there would be no reason to look further for a natural explanation. Use of supernatural explanations would shut down science. … If science relied on a creator to explain the inexplicable, there would be nowhere to go, no way to prove that explanation wrong. The question would have already been settled. … Science does not deny the existence of God. God is simply outside its realm.[57]

---